\pdfoutput=1

\documentclass[twocolumn,showpacs,amssymb,aps,nofootinbib,floatfix,superscriptaddress]{revtex4-1}

\usepackage{epsfig}
\usepackage{hyperref}
\usepackage{comment}

\usepackage{graphicx,color}

\begin{document} 
\hbadness=10000

\title{$\alpha$
clusters and collective flow in ultra-relativistic carbon--heavy 
nucleus collisions}

\author{Piotr Bo\.zek}
\email{Piotr.Bozek@ifj.edu.pl}
\affiliation{AGH University of Science and Technology, Faculty of Physics and
Applied Computer Science, al. Mickiewicza 30, 30-059 Krakow, Poland}
\affiliation{The H. Niewodnicza\'nski Institute of Nuclear Physics, Polish
Academy of Sciences, PL-31342 Cracow, Poland}

\author{Wojciech Broniowski}
\email{Wojciech.Broniowski@ifj.edu.pl}
\affiliation{Institute of Physics, Jan Kochanowski University, 25-406 Kielce, Poland}
\affiliation{The H. Niewodnicza\'nski Institute of Nuclear Physics, Polish
Academy of Sciences, PL-31342 Cracow, Poland}

\author{Enrique Ruiz Arriola}
\email{earriola@ugr.es}
\affiliation{Departamento de F\'{\i}sica At\'{o}mica, Molecular y Nuclear and Instituto Carlos I de  F{\'\i}sica Te\'orica y Computacional, 
 Universidad de Granada, E-18071 Granada, Spain}

\author{Maciej Rybczy\'nski}
\email{Maciej.Rybczynski@ujk.edu.pl}
\affiliation{Institute of Physics, Jan Kochanowski University, 25-406 Kielce, 
Poland}

\begin{abstract}
We investigate ultrarelativistic collisions of the $^{12}$C nucleus 
with heavy targets and show that the harmonic flow measures based on ratios of
cumulant moments are particularly suited to study the intrinsic deformation of 
the $^{12}$C wave function. That way one can probe the 
expected $\alpha$ clusterization in the ground state, which leads to large 
initial triangularity in the shape of the fireball in the transverse plane. 
We show that the clusterization effect results in very characteristic behavior 
of the ratios of the cumulant moments as functions of the number of participant 
nucleons, both for the elliptic and triangular deformations. 
Thus the experimental event-by-event studies of harmonic flow in 
ultrarelativistic light-heavy
collisions may offer a new window to look at the ground-state 
structure of light nuclei.
\end{abstract}

\pacs{21.60.Gx, 25.75.Ld}

\maketitle

\section{Introduction \label{sec:intro}}

In a recent paper~\cite{Broniowski:2013dia} a novel method of
investigating the $\alpha$ clusterization of light nuclei was
proposed, exploring an unexpected bridge between the lowest-energy
nuclear physics determining the ground-state structure and the
highest-energy nuclear collisions.
In this work we further pursue the method of  
searching for specific signals of the intrinsic geometry of light
nuclei manifest in harmonic flow. 
Our approach is based on the fact that at the ultra-high collision energies, 
where
the nucleon-nucleon inelastic collisions generate a stream of copious particles,
the interaction times are short enough to prevent the much slower
nuclear excitations.  Therefore, the initial spatial distributions of
nucleons in the nuclear ground state of the overlaying nuclei
mark the location of sources (inelastic collisions) igniting the
fireball. 

Atomic nuclei have a genuine degree of granularity which can be
characterized by typical correlation lengths and which reflects the
energetically favored spatial orderings. Any such geometric feature
should leave some fingerprint in the final state, provided these
effects and the random fluctuations due to finite number of particles can be
cleanly separated. The theoretical identification and the experimental
verification of these geometry-preserving features would provide not
only valuable information on conventional nuclear structure, where
genuine multi-nucleon aspects of the nuclear wave functions could be
tested, but would also generate confidence on the currently intricate
theoretical protocols and the approximation schemes used to analyze
the dynamics of relativistic heavy-ion collisions.
 
In this paper we analyze in further
detail the possible experimental signatures of presence of the
$\alpha$ clusters in $^{12}$C, which is of direct significance for
analysis of its ultrarelativistic collisions with a heavy target. In
particular, we focus on the ratios of harmonic flow measures
investigated in the hydrodynamic framework~\cite{Bozek:2014cya} for the case of the $^3$He-Au
collisions and defined in a suitable way such that the sensitivity to
the details of the intermediate dynamical/collective stages of the
fireball evolution is eliminated. Such careful procedures are needed,
as the studied observables carry information on both the initial
geometry of the fireball, as well as on its random fluctuations which
have the tendency of covering up the geometry to some extent.

We show that the intrinsic triangular geometry of $^{12}$C,
predetermined by the arrangement of the $\alpha$ clusters, leads to
very specific and pronounced dependence of the considered measures on
the number of nucleons participating in the collision. We argue that
the proposed method provides practical tools to investigate signatures
of the cluster structure of the ground-state wave function of light
nuclei, and that it could be directly employed in studies of future
ultrarelativistic heavy-ion collisions of light-heavy systems.

The paper, designed for both the researchers of the
$\alpha$-clusterization and the relativistic heavy-ion community, is
arranged as follows: In Sec.~\ref{sec:over} we briefly review the
current status of the $\alpha$ clusterization in light nuclei, in the
scope needed for our work. Based on that knowledge we prepare our
nuclear Monte Carlo configurations of $^{12}$C as described in
Sec.~\ref{sec:genc}. These configurations are later used in the
simulations of collisions with a heavy target.

Then we pass to presenting our modeling of the early stage of the
collision, stressing its quantum-mechanical aspects in
Sec.~\ref{sec:quant}. The important point here is that the reaction
time at ultrarelativistic energies is much shorter from any typical
nuclear-dynamics time, resulting in a frozen configuration of the
nucleons reflecting the structure of the ground-state nuclear wave
function.  The formation of the fireball is described in
Sec.~\ref{sec:fire}, where we apply the popular Glauber
approach~\cite{Glauber:1959aa,Czyz:1969jg,Bialas:1976ed,Miller:2007ri,%
Broniowski:2007ft,Bialas:2008zza} for our event-by-event studies. The
key quantities here are the eccentricity parameters, defined in
Sec.~\ref{sec:eccent}.

In Sec.~\ref{sec:flow} we turn to harmonic flow -- phenomenon used 
extensively in the relativistic heavy-ion programs to infer properties of 
the dynamically evolving quark-gluon plasma with the help of well-developed 
methods~\cite{Ollitrault:1992bk,Borghini:2001vi,%
Voloshin:2002wa,Voloshin:2008dg}. From our point of view, the essential feature 
here 
is the approximately linear response of the dynamical system to the eccentric 
deformation of the initial state, resulting in proportionality of the measurable 
flow coefficients to the corresponding initial eccentricity parameters. The 
unknown response coefficient may be eliminated by taking appropriate ratios of 
moments of the distribution, as explained in Sec.~\ref{sec:ratios}. In 
particular, we consider moments based on the two- and four-particle cumulants, 
used frequently in experimental studies. Such ratios for the flow moments are 
equal to the corresponding ratios of the eccentricity moments, thus allowing for 
predictions 
of the measured quantities related to flow based solely on measures of the 
initial state. Moreover, these ratios are sensitive to the geometry and random 
fluctuations in a very specific way. 

In particular, we find that for the $^{12}$C collisions on a heavy target, 
the ratio of the four- to two-particle cumulant moments changes behavior for 
the high-multiplicity events (centrality below 10\%), increasing with the 
number of participating nucleons for triangularity, and decreasing for
ellipticity, in accordance with the geometric features of the system. This is 
the key result of this work.

The feature holds at various collision energies 
(Sec.~\ref{sec:energy}) and rapidities (Sec.~\ref{sec:rapidity}), as well as 
for different Glauber models of the initial state (Sec.~\ref{sec:deformed}. 
We have also checked the
dependence of our results on the model of the $^{12}$C wave function 
(Sec.~\ref{sec:deformed}).

\section{$\alpha$ clusters in light nuclei \label{sec:alpha}}

\subsection{Overview \label{sec:over}}

The idea of $\alpha$ clustering dates back to the old work of
Gamow~\cite{gamow1931constitution}, where he conceived the radioactive
$\alpha$-decay process as a signal for nuclear constituents. This is
in agreement with the tight binding ($B_\alpha/4 \sim 7~{\rm MeV}$), and
compactness ($r_\alpha \sim 1.5 {\rm fm}$) of the quartet of states
($p \uparrow$, $p \downarrow$, $n \uparrow$, $n \downarrow$) building
the $^4$He nucleus and the weak $\alpha\alpha$ attraction
($V_{\alpha\alpha} \sim -2.5 {\rm MeV}$) gluing the $\alpha$-particles
into the light $A=4n$
nucleus~\cite{PhysRev.52.1083,PhysRev.54.681,wefelmeier1937geometrisches}
(see, e.g.,~\cite{blatt19521952theoretical} for an early review
and~\cite{brink2008history} for a historic account). This has been
one of the most fascinating issues of the nuclear structure physics
throughout decades (for reviews of the topic see,
e.g.,~\cite{brink1965alpha,freer2007clustered,%
ikeda2010clusters,beck2012clusters,%
Okolowicz:2012kv,Zarubin,Beck:2014fja}).

The $^{12}$C nucleus is of particular interest. It was
described as a bound state of three elementary $\alpha-$particles by
Harrington~\cite{PhysRev.147.685} (for a review and further
references see~\cite{Freer:2014qoa}). Numerous theoretical
approaches were applied to its ground-state and excited states structure:
the Bose-Einstein Condensation (BEC) model~\cite{Funaki:2006gt}, the
fermionic molecular dynamics (FMD)~\cite{Chernykh:2007zz},
antisymmetrized molecular dynamics~\cite{KanadaEn'yo:2006ze},
effective chiral field theory on the lattice~\cite{Epelbaum:2012qn},
the no-core shell model~\cite{PhysRevLett.109.052501,Barrett:2013nh},
or the variational Green's function method
(VMC)~\cite{Pieper:2002ne}.  The recently discovered $5^-$
rotational state of $^{12}$C in low energy $\alpha+^{12}$C collisions
points to the triangular ${\cal D}_{3h}$ symmetry of the
system~\cite{Marin-Lambarri:2014zxa}.

\subsection{Generating clustered distributions \label{sec:genc}}

Our aim is to consider $\alpha$-clustered ground-state structure of
$^{12}$C.  Ideally, distributions following from realistic model wave
functions or ab initio calculations should be used. As this is quite
involved, or requires access to the Monte Carlo nuclear configurations
in the path-integral Green's function methods, we proceed in a
simplified manner which grasps the essential features of the
ground-state distributions and serves our purpose to sufficient
accuracy. As we wish to study the effects of clusterization, we assume
that $^{12}$C is formed of three separated $\alpha$ clusters. The
parameters of the arrangement are adjusted in such a way that the
desired one-body radial density of the centers of nucleons is
reproduced, as described below.

\begin{figure}[tb]
\centering
\includegraphics[angle=0,width=0.37 \textwidth]{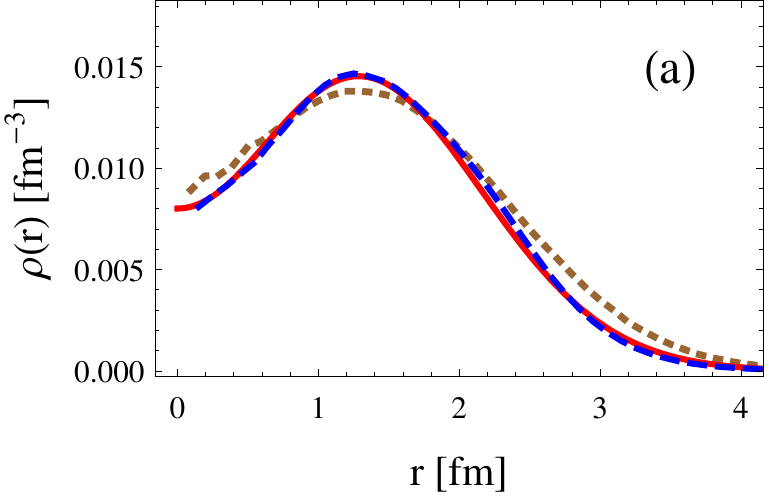} \\ 
\vspace{6mm}
\includegraphics[angle=0,width=0.37 \textwidth]{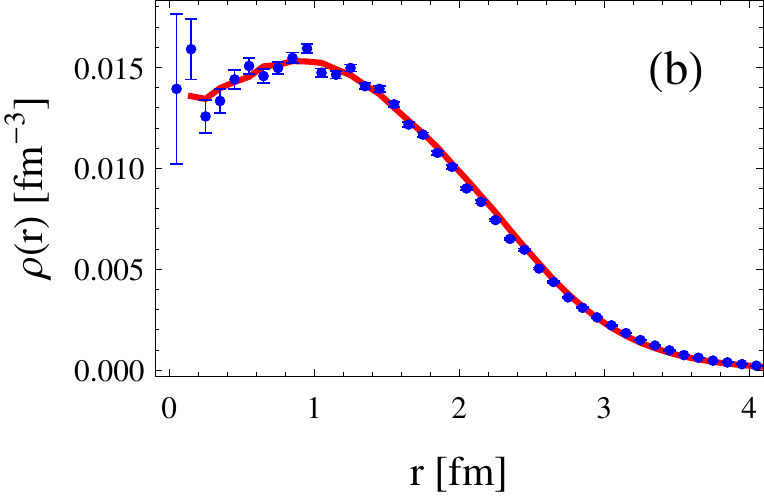}
\vspace{-3mm}
\caption{(Color online) \label{fig:dense} (a) Radial distribution of the 
centers of nucleons in $^{12}C$ for the BEC calculation reproducing the charge 
form factor (dashed line), our parametrization of the BEC calculation (solid 
line), and the Jastrow calculation of Ref.~\cite{buendia2001projected} (dotted 
line). (b)~The VMC calculation (points) and our parametrization (solid line). 
See text for details.}
\end{figure}

Technically, we carry out the Monte Carlo simulations of the nuclear
configurations as follows: we place the centers of the three clusters
in an equilateral triangle of edge length $l$. The distribution in
each cluster is a Gaussian parametrized as
\begin{eqnarray}
f_i(\vec{r})=A \exp \left (- \frac{3}{2} \, (\vec{r}-\vec{c_i})^2/r_\alpha^2 
\right ),  
\end{eqnarray}
where $\vec{r}$ is the coordinate of the nucleon, $\vec{c_i}$ is the
position of the center of the cluster $i$, and $r_\alpha$ is the rms
radius of the cluster. We generate randomly the positions of the
nucleons, in sequence alternating the number of the cluster: 1, 2, 3,
1, 2, 3, etc., and taking into account the short-distance NN repulsion
in a popular way, where the centers of each pair of nucleons cannot be
closer than 0.9~fm~\cite{Broniowski:2010jd}.  At the end of the
procedure the distributions are shifted such that their center of mass
is placed at the origin of the coordinate frame.  As a result, we get
the Monte Carlo $^{12}$C distributions with the built-in
$\alpha$-cluster correlations.

The model parameters $l$ and $r_\alpha$ are optimized in such a way
that the desired form of the radial density is obtained.  Thus the
radial density of the centers on nucleons serves as a constraint for
building the clustered distributions.  Throughout this paper we use
two reference radial distributions: those obtained from the so-called
Bose-Einstein Condensation (BEC) model~\cite{Funaki:2006gt} and
distributions from the variational Monte Carlo calculations (VMC)
using the Argonne~v18 two-nucleon and Urbana~X three-nucleon
potentials, as recently provided in
{\small \url{http://www.phy.anl.gov/theory/research/density}}.
Figure~\ref{fig:dense} shows the quality of our fit to the one-body
densities for the two considered cases: BEC in Fig.~\ref{fig:dense}(a)
and VMC in Fig.~\ref{fig:dense}(b), where we also give for a reference
the result of the Jastrow-type correlated wave
function~\cite{Buendia:2004yt}.

\begin{table}[tb]
\caption{Parameters used in our Monte Carlo simulations for the distributions 
of nucleons in $^{12}$C. \label{tab:param}}
\vspace{3mm}
\begin{tabular}{|c|rr|}
\hline
parameter & BEC & VMC \\
\hline
$l$ [fm]        & 3.05  & 2.84   \\
$r_\alpha$ [fm] & 0.96  & 1.15   \\
\hline 
\end{tabular}
\end{table}

The parameters used in our simulations are collected in
Table~\ref{tab:param}. As $l$ is much larger than $r_\alpha$, the
distributions are hollow in the center, and the curves in
Fig.~\ref{fig:dense} exhibit a dip, more depleted as
$r_\alpha/l$ decreases.  We note that, after properly
implementing the nucleon charge contribution with $r_N=0.87 {\rm fm}$, the BEC 
densities reproduce very well the charge form factor of
$^{12}$C, which is not the case of VMC, albeit the charge density near
the origin carries rather large uncertainties inherited from
the lack of knowledge in the high-momentum region in the
measured charge form factor. As the BEC case is more strongly
clustered than the VMC case ($r_\alpha/l$ is smaller), we shall use it
as our basic model to illustrate the investigated effects, which are
stronger with this choice. We will occasionally compare also to the
VMC scenario.

As we are interested in specific effects of clusterization,
as a ``null result'' we also use the {\em uniform} distributions,
i.e., with no $\alpha$ clusters.  We prepare such distributions with
exactly the same radial density as the clustered ones. This is
achieved easily with a trick, where we randomly regenerate the
spherical angles of the nucleons from the clustered distributions, while 
leaving the
radial coordinate intact.

While the above-described procedure may seem crude for the description
of the nuclear structure, it is sufficient for our goal. We note that
the method reproduces not only the one-body densities, as shown above,
but also the two-particle densities determined from multicluster
models with the state-dependent Jastrow
correlations~\cite{Buendia:2004yt} are describes with a reasonable
accuracy ($\sim10-20\%$)
\cite{Broniowski:2014aqa} . 

\section{Early-stage of the ultra-relativistic reaction \label{sec:early}}

\begin{figure*}[tb]
\centering
\includegraphics[angle=0,width=0.32 \textwidth]{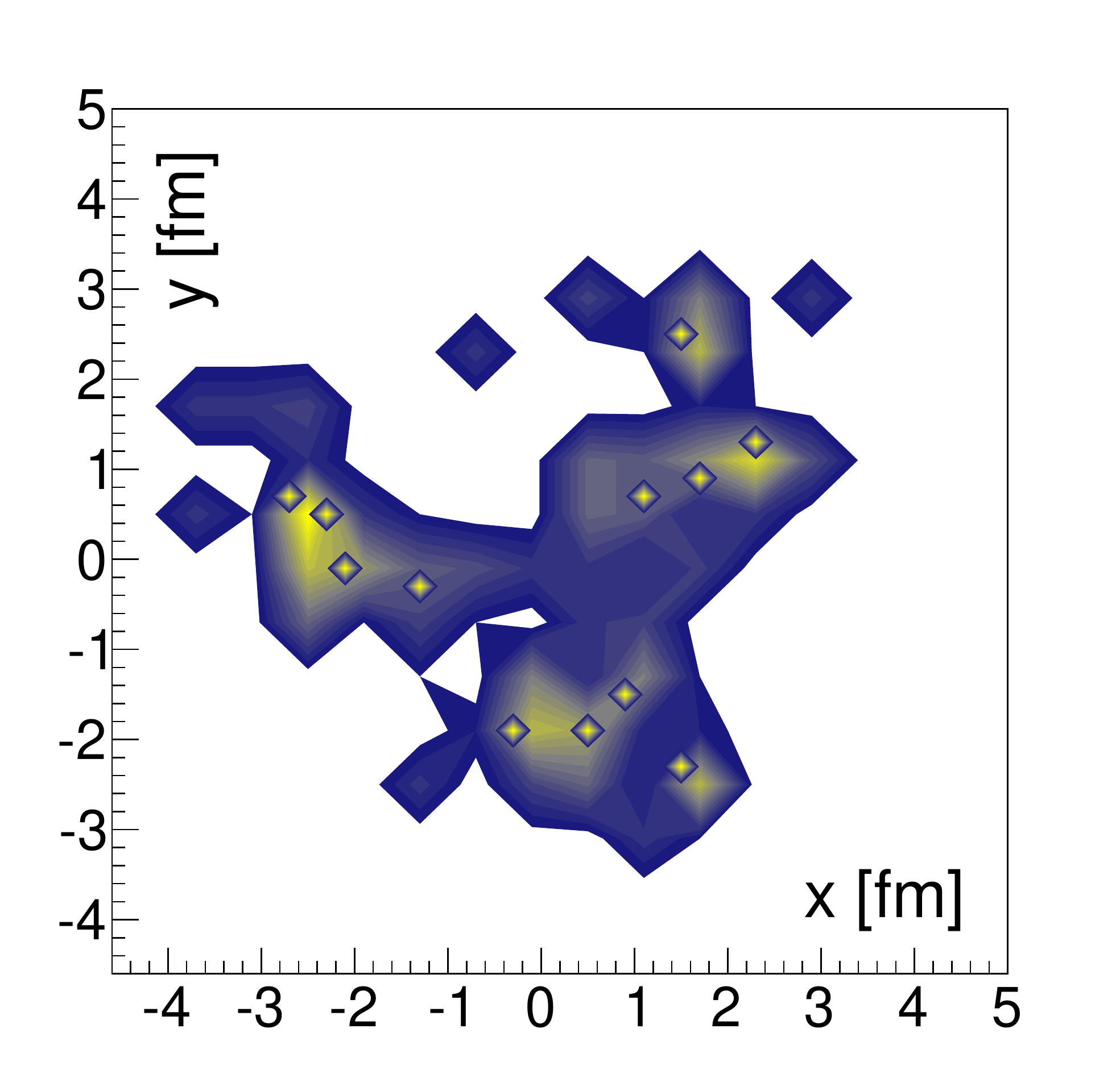}
\includegraphics[angle=0,width=0.32 \textwidth]{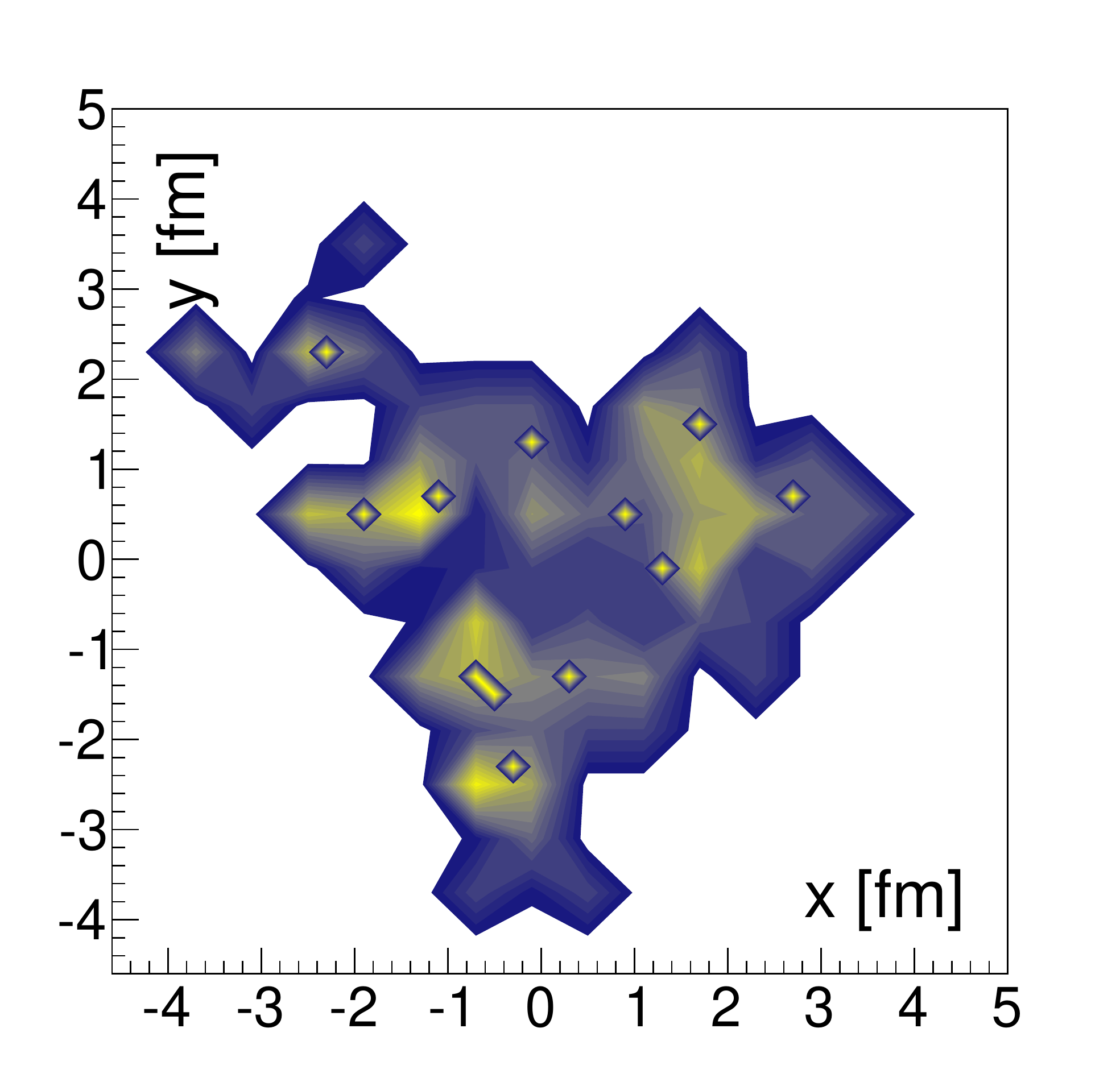}
\includegraphics[angle=0,width=0.32 \textwidth]{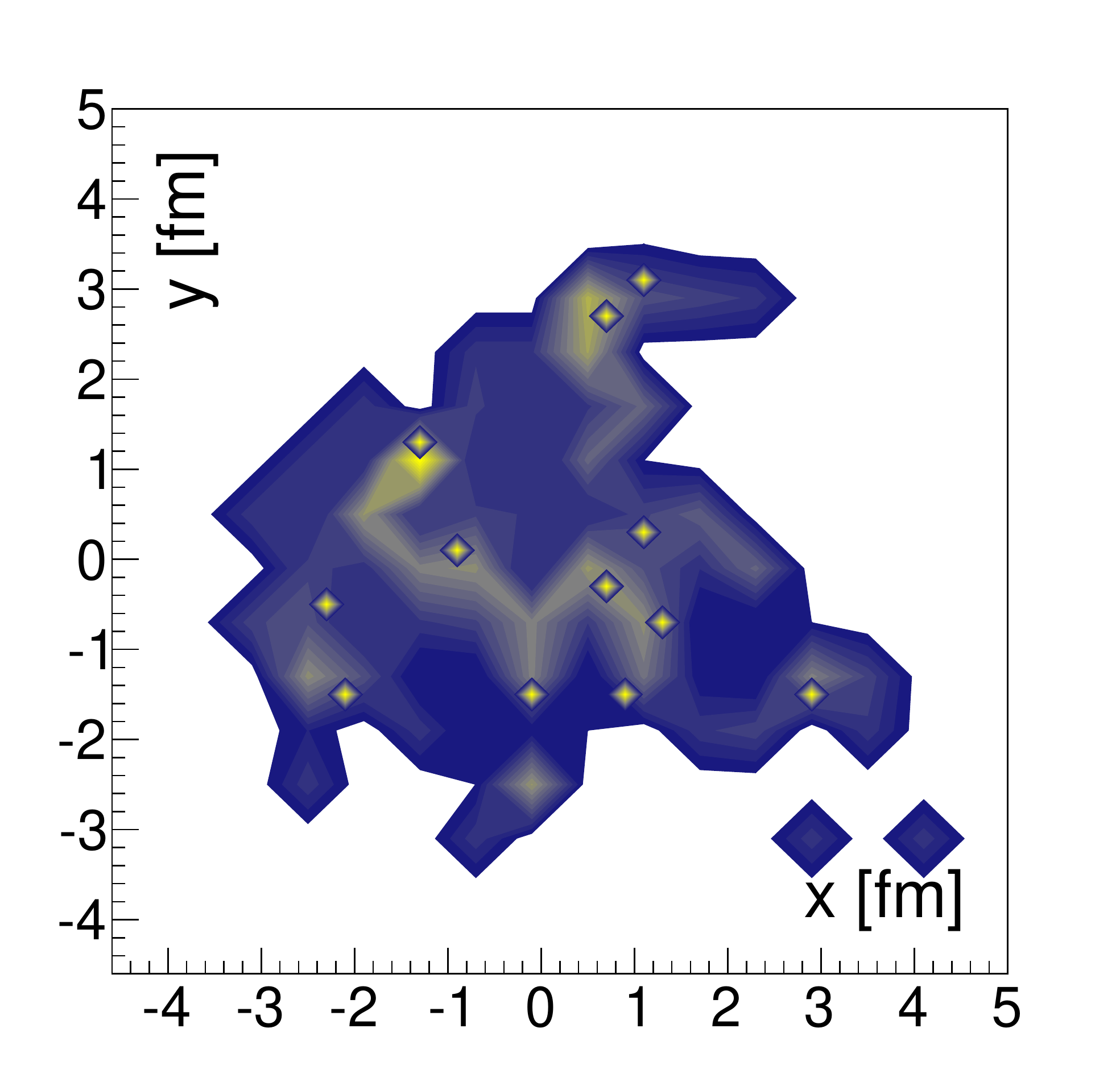}
\vspace{-3mm}
\caption{(Color online) \label{fig:fire} Snapshots of three sample
${}^{12}$C--${}^{197}$Au events, displaying 
the distribution of sources in the transverse plane. The small diamonds indicate
the positions of the $^{12}$C nucleons, while the dark region shows the density 
of the fireball including the wounded nucleons from $^{197}$Au and the binary 
collisions. BEC case, RHIC, $N_{\rm w}=66$. 
In this simulation the transverse 
and cluster planes were aligned for better visualization. See text for details.}
\end{figure*}

\subsection{Quantum-mechanical aspects of the collision \label{sec:quant}}

Viewed in the lab frame of collider experiments, the colliding
ultra-relativistic nuclei move almost with the speed of light. The
corresponding Lorentz contraction factors are very large ($\sim$1000
at the LHC and $\sim$100 at the RHIC energies), such that we deal
with collisions of ``flat pancakes'' and hence the initial-state
interactions are negligible; the reaction time is much shorter than
any typical and slow nuclear structure time scale to witness any
relevant nuclear excitation. As a result, in the reaction, a frozen
nuclear ground-state configuration is seen.  The wave function
undergoes quantum-mechanical reduction in the earliest stage of the
reaction, which results in a given intrinsic nuclear
configuration. As outlined above, we generate event-by-event
probability following the square of the nucleus wave function.

The heavy nucleus which collides with $^{12}$C (here $^{197}$Au or
${}^{208}$Pb) is made according to the Monte Carlo procedure described
in detail~\cite{Broniowski:2007nz,Rybczynski:2013yba}. The
short-distance NN repulsion and the nuclear deformation effects are
taken into account. 

Note that since $^{12}$C is much smaller than
$^{197}$Au or ${}^{208}$Pb, for sufficiently small impact parameters the 
collisions correspond to the $\alpha$-clustered, triangle-shaped, 
and randomly oriented nucleus bumping into the central region of a heavy
nucleus. This can pictorially be idealized as an equilateral triangle
of three-$\alpha$'s {\em hitting a wall} of nuclear matter.

In the following we are concerned with the transverse plane, relevant
for the mid-rapidity physics studied later, hence we only need the
wave functions in the transverse plane and mid-rapidity. The typical
locations of the centers of $^{12}$C nucleons are indicated with small
diamonds in Fig.~\ref{fig:fire}.

\subsection{Formation of the fireball \label{sec:fire}}

The initial density of the fireball in ultra-relativistic heavy-ion
collisions is formed out of the individual NN collisions between
nucleons from the two colliding nuclei. At these high energies
most collisions are inelastic and copious particles (partons) are produced.
Popular modeling of this phase is accomplished with the Glauber
approach
\cite{Glauber:1959aa,Czyz:1969jg,Bialas:1976ed,Miller:2007ri,%
Broniowski:2007ft,Bialas:2008zza}, 
applied in this 
work. One may alternatively 
adopt the 
Kharzeev-Levin-Nardi framework
\cite{Kharzeev:2000ph,Kharzeev:2002ei,Drescher:2006pi} based on the Color Glass 
Condensate model which rests explicitly on quark-gluon dynamics
(see, e.g.,~\cite{Albacete:2014fwa}).

Within the Glauber framework, we use the so-called mixed
model~\cite{Kharzeev:2000ph,Back:2001xy}, where the entropy deposition
in the transverse plane comes from the wounded
nucleons~\cite{Bialas:1976ed}, defined as those who interacted
inelastically at least once, and from the binary collisions. A source
coming from the wounded nucleon is placed in its center with a
relative weight $(1-a)/2$, while the location of the binary collision
is in the center of mass of the colliding nucleon pair, while the
relative weight is $a$.  The probability of the collision
is defined relative to the total inelastic NN cross section,
$\sigma_{\rm NN}^{\rm inel}$, corresponding to the CM collision energy
in the process.
The NN collision profile corresponds to the probability of
an NN inelastic collision to happen at a given impact parameter
$b$. We use the smooth function described
in~\cite{Broniowski:2007nz,Rybczynski:2013yba}, which is constructed
to reproduce approximately the differential elastic NN cross section. The 
values of
parameters used in our simulations are listed in Table~\ref{tab:par}. 

\begin{table}[tb]
\caption{\label{tab:par} Parameters used in the GLISSANDO simulations.}
\vspace{3mm}
\begin{tabular}{|c|crcl|}
\hline
 & system & $\sqrt{s_{\rm NN}}$ [GeV] & $\sigma_{\rm NN}^{\rm inel}$ [mb] & $a$ 
\\
\hline 
SPS  & $^{12}$C+$^{208}$Pb & 17   & 32 & 0.12 \\
RHIC & $^{12}$C+$^{197}$Au & 200  & 42 & 0.145 \\
LHC  & $^{12}$C+$^{208}$Pb & 5200 & 73 & 0.15 \\
\hline
\end{tabular}
\end{table}


The Monte Carlo simulation produces, in each
event, the location of the centers of the sources distributed in the
transverse plane. Physically, sources generated in a collision process
are of non-zero width, reflecting the production mechanism (non-zero
size of nucleons, flux tubes, etc.) and thus the sources must be {\em
smeared}. This feature can be modeled by placing a two-dimensional
Gaussian of width $\sigma$ (in this paper we take $\sigma=0.4$~fm)
centered at the source in the transverse plane. This {\em physical}
smearing effect is necessary in forming the initial condition for
hydrodynamics, taking over the evolution of the system. Also, smearing
is phenomenologically important for the shape eccentricities, which
are significantly reduced compared to the naive evaluation with
point-like sources. This smearing length sets a typical coarse
graining scale for hydrodynamics beyond which the integration step
would not explore the relevant physics.

Sample events of central
${}^{12}$C--${}^{197}$Au collision are presented in Fig.~\ref{fig:fire}. 
We have used here the 
clustered ${}^{12}$C BEC distributions and, for the purpose of better 
visibility, aligned the transverse and the cluster planes (the carbon hits the 
lead ``flat''). 
The small diamonds mark
the positions of the $^{12}$C nucleons, while the dark region represents the 
transverse density 
of the fireball, including the wounded nucleons from $^{197}$Au and the binary 
collisions. The irregular ``warped'' structure follows from the stochastic 
nature of the process. Nevertheless, one may easily notice the remnant 
triangular shape in the fireball distribution, originating from the underlying 
three $\alpha$ clusters in $^{12}$C.

\subsection{Eccentricities of the initial state \label{sec:eccent}}

With the smeared sources, the eccentricity parameters of the 
fireball, $\epsilon_n$, are defined in each event via the Fourier decomposition 
of the density in the transverse plane,
\begin{eqnarray}
\epsilon_n e^{i n \Phi_n} = - \frac{\int dx dy \, f(x,y) \rho^n e^{i n 
\phi}}{\int dx dy \, f(x,y)  \rho^n}, \label{eq:epsdef}
\end{eqnarray}
where $n=2,3,\dots$ is the rank, $\Phi_n$ is the angle of the principal axes, 
$x$ and $y$ are coordinates in the transverse plane, with 
$\rho=\sqrt{x^2+y^2}$ and ${\rm tan} \phi = y/x$, finally, $f(x,y)$ is the 
fireball density in the given event.
The $n=2$ eccentricity is termed {\em ellipticity}, and $n=3$ {\em 
triangularity}. 

Non-vanishing contributions to the coefficients $\epsilon_n$ come from two 
different origins. One of them is
 the intrinsic ``geometry'' of the distribution of the nucleons in
${}^{12}$C. In the clustered case there is a large triangularity of this
distribution from the arrangement of the $\alpha$ clusters in an
equilateral triangle. Although it is somewhat reduced due to random
orientation of the ${}^{12}$C nucleus with respect to the transverse
plane, the values of $\epsilon_3$ remain sizable. The ${}^{12}$C
nucleus also exhibits geometric ellipticity in the case when the
cluster plane is not parallel to the transverse plane, which is the
generic case due to randomness of the orientation.

\begin{figure}
\centering
\includegraphics[angle=0,width=0.46 \textwidth]{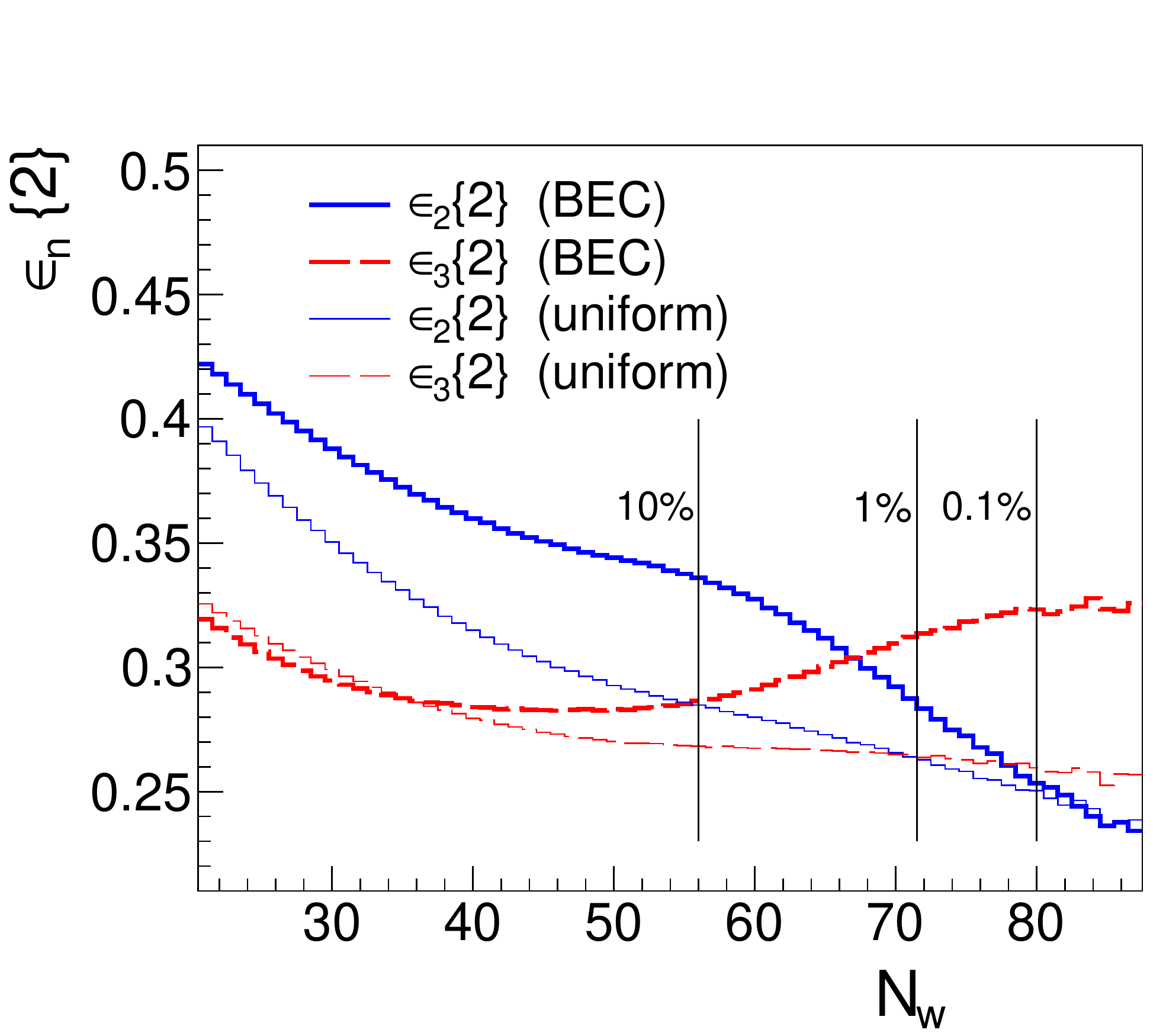}
\vspace{-3mm} 
\caption{(Color online) Comparison of the fireball eccentricity coefficients 
from the two-particle cumulants for the clustered distribution and for the 
uniform distribution. GLISSANDO simulations, BEC case, RHIC. The vertical lines 
indicate total number of the wounded nucleons corresponding to centralities 
10\%, 1\% and 0.1\%. The orientation-multiplicity correlation is clearly seen 
for the clustered case.
\label{fig:eps23}}
\end{figure}
\begin{figure}
\centering
\includegraphics[angle=0,width=0.42 \textwidth]{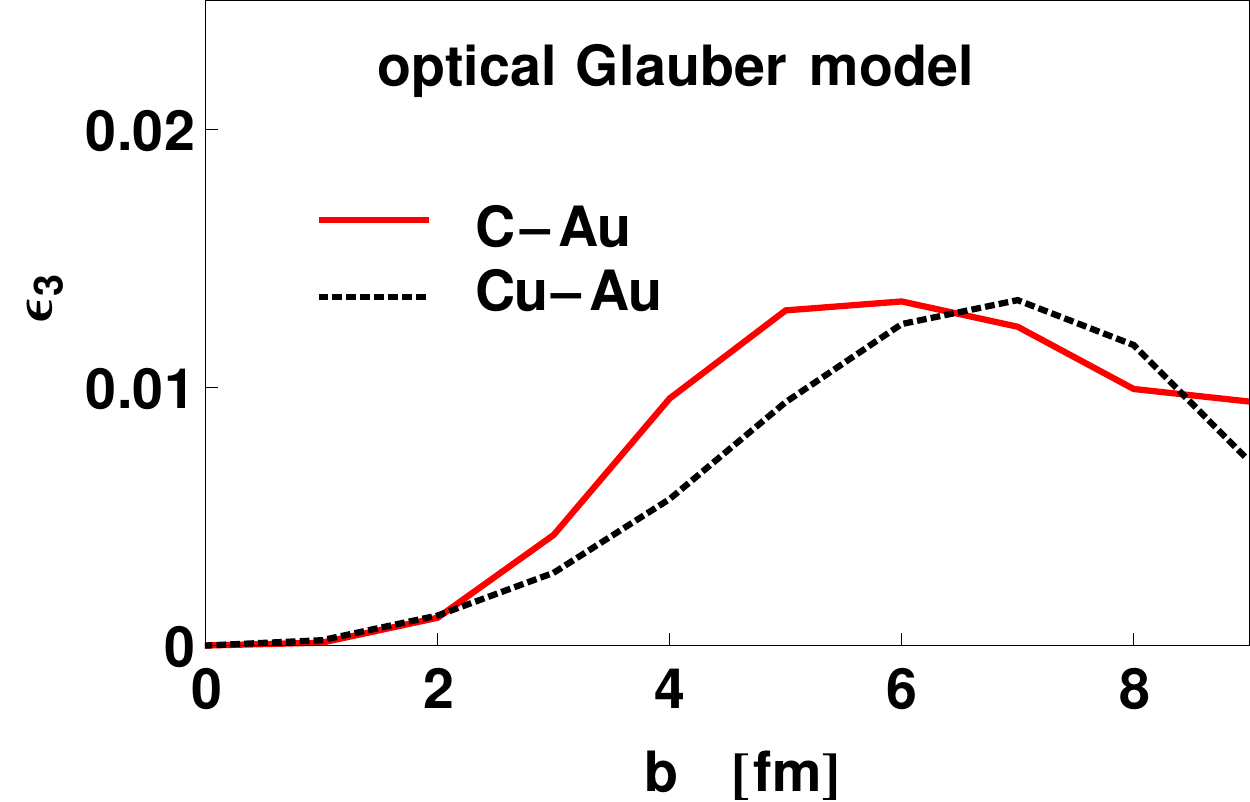}
\vspace{-3mm} 
\caption{(Color online)  Triangularity of the fireball formed in C-Au and Cu-Au collisions. The density is calculated in the optical Glauber model, and the the triangularity is defined with respect to the reaction plane.
\label{fig:ope3}}
\end{figure}

The second cause for eccentricity coefficients comes from fluctuations
due to the finite number of
nucleons~\cite{Miller:2003kd,Manly:2005zy,Voloshin:2006gz,Broniowski:2007ft,%
Alver:2010gr}. The effect of fluctuations washes away to some extent
the geometric component, hence a careful examination of the results
presented in Sec.~\ref{sec:ratios} is necessary to discriminate the
two different origins.

In collisions of asymmetric nuclei at a finite impact parameter $b$, 
small values of odd Fourier components can appear in the azimuthal
dependence of the fireball density with respect to the reaction plane
(such an effect is present, for instance, in the Cu-Au collisions).
In Fig.~\ref{fig:ope3} we show the triangularity of the initial
fireball for the C-Au and Cu-Au collisions with respect to the
reaction plane calculated in the optical Glauber model, an
approximate scheme where one first averages the densities and then
computes the nuclear thickness function~\cite{Florkowski:2010zz}. For
intermediate values of $b$ the triangularity is non-zero, even without
any contribution from fluctuations or the $\alpha$
clustering. However, the obtained value of $\epsilon_3$ is an order of
magnitude smaller than the one calculated event-by-event with respect
to the third order event plane (Fig. \ref{fig:eps23}). Moreover, the
most central collisions that we discuss in the following correspond to
small impact parameters (for centralities $c=10\%$, 1\% and 0.1\% the
average values of $b$ are 2.4, 1.5, and 1.2~fm, respectively).  Hence
the average geometric $\epsilon_3$ in the reaction plane is even
smaller.  While the above effect is automatically included in our simulation, it
does not play a role in the interpretation of the results.

\begin{figure}
\centering
\includegraphics[angle=0,width=0.3 \textwidth]{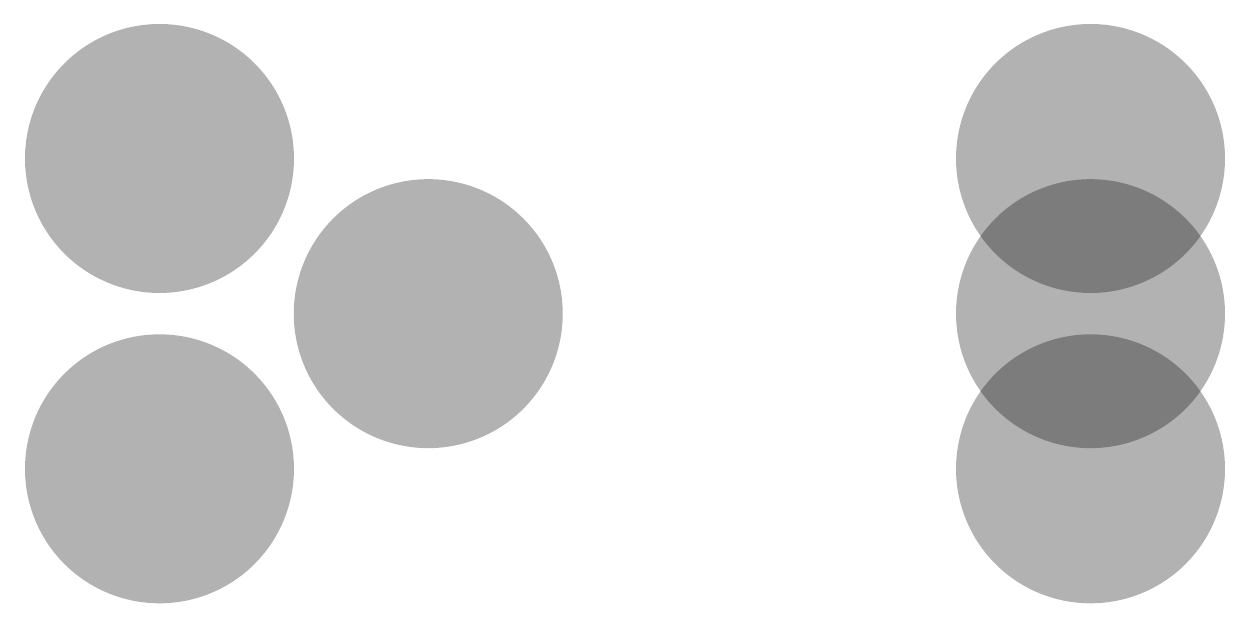}
\vspace{-3mm} 
\caption{(Color online) The flat-on (left) and side-wise (right) orientations 
of $^{12}$C with respect to the reaction plane. 
\label{fig:flat}}
\end{figure}

As already explained in~\cite{Broniowski:2013dia}, there is a specific 
correlation between centrality, triangularity, and ellipticity, induced by the 
intrinsic orientation of $^{12}$C.
When the transverse and the cluster planes are aligned, the $^{12}$C nucleus 
hits the large nucleus flat-on and thus creates most damage, i.e., produces the 
largest number of sources (cf. left side of Fig.~\ref{fig:flat}). At the same 
time, in this flat-on orientation we 
have on the average the highest triangularity 
and the lowest ellipticity, which here comes entirely from fluctuations. 

In the other extreme case the cluster plane is perpendicular to the 
transverse plane (side-wise configuration, cf. right side of 
Fig.~\ref{fig:flat}). Then we find the opposite behavior: 
low multiplicity, as the cross section is smaller, small triangularity,
and large ellipticity, which now obtains a sizable contribution from the 
elongated shape of the fireball. 

Of course, in actual collisions the orientation is random and we have a 
situation between the two limiting cases described above, yet the phenomenon of 
the specific orientation-multiplicity correlations is clearly seen (cf. 
top-right Fig.~3 of Ref.~\cite{Broniowski:2013dia} or Fig.~\ref{fig:eps23} 
in this paper). In particular, in Fig.~\ref{fig:eps23} we see, by comparing the 
simulations with clustered and uniform $^{12}$C, that the geometry raises 
triangularity at high values of the number of wounded nucleons, $N_w$ 
(preferentially flat-on collisions), and raises ellipticity at lower values of 
$N_w$ (side-wise collisions).

Event-by-event studies allow for obtaining event-by-event distributions of the 
physical quantities. In the Sections below we will need the so-called 
two-particle and four-particle cumulants moments~\cite{Borghini:2000sa} of the 
eccentricities, defined as
\begin{eqnarray}
     \epsilon_n\{2\} &=&  \langle \epsilon_2^2 \rangle^{1/2}, 
\nonumber \\
     \epsilon_n\{4\} &=& 2 \left ( \langle \epsilon_n^2 \rangle^2 -
\langle \epsilon_n^4 \rangle \right )^{1/4}. \label{eq:cumul}
\end{eqnarray}
For a finite number of sources (wounded nucleons), even without geometric 
deformation, one has just from fluctuations $\epsilon_n\{m\}\neq 0$ for $m \ge 
4$, with $\epsilon_n\{m\}$ decreasing as 
$1/N_{w}^{1-1/m}$~\cite{Bhalerao:2006tp,Alver:2010gr}.

\section{Collectivity and development of harmonic flow \label{sec:harm}}

\begin{figure}
\centering
\includegraphics[angle=0,width=0.45 \textwidth]{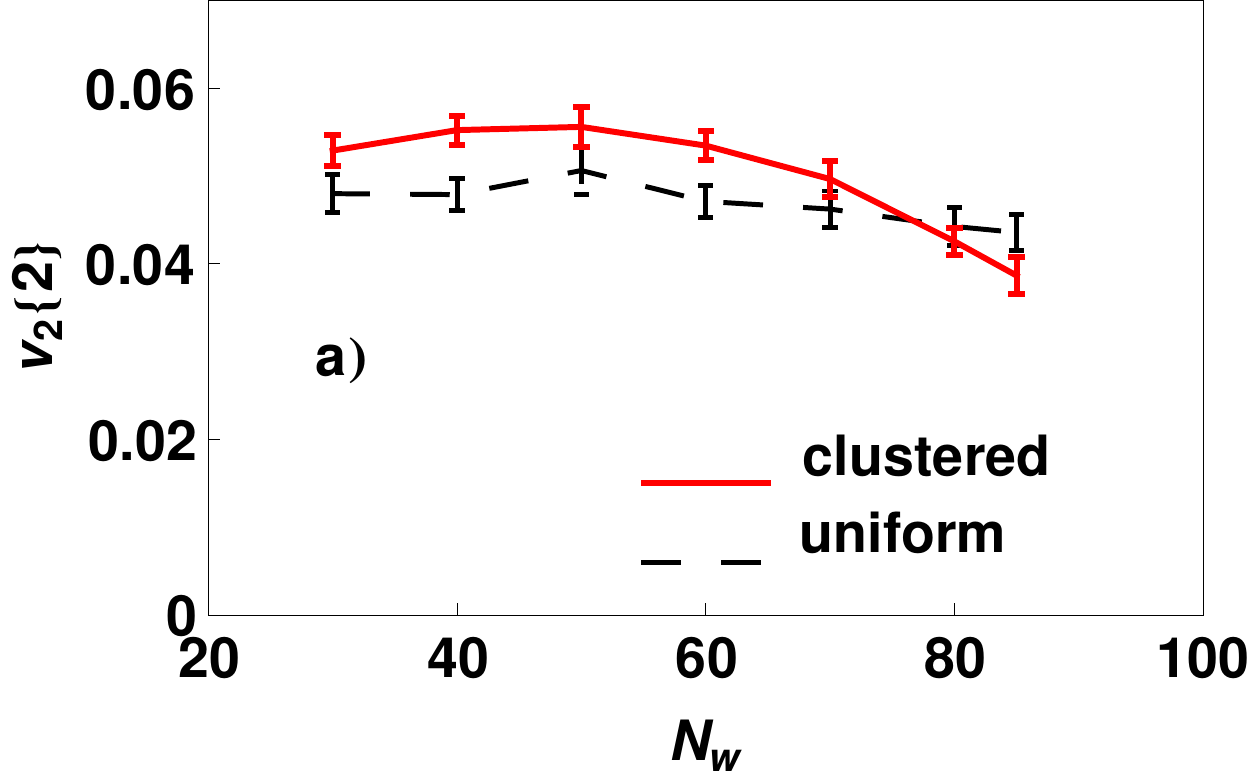}
\includegraphics[angle=0,width=0.45 \textwidth]{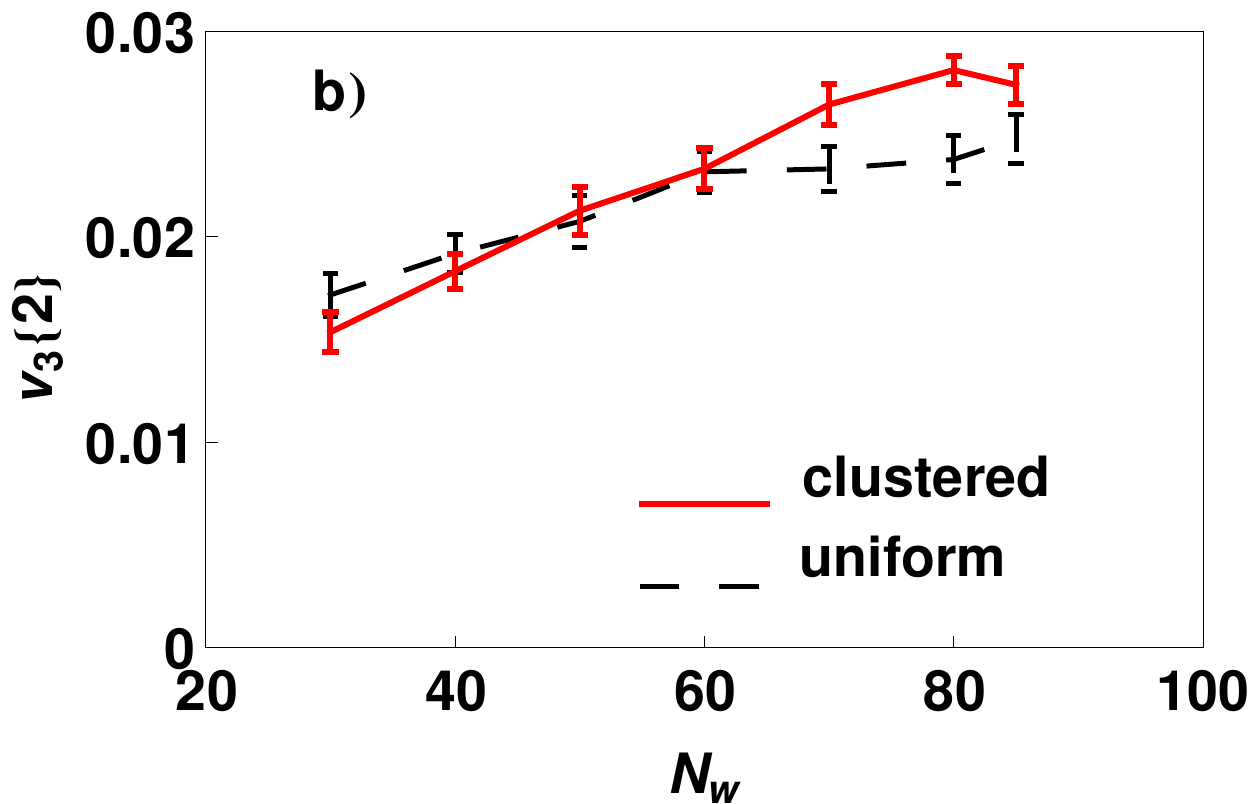}
\vspace{-3mm} 
\caption{(Color online) Elliptic (panel a) and triangular (panel b) 
flow coefficients as a function of the number of wounded nucleons for  C-Au collisions, calculated using
 the second order cumulant. \label{fig:v2v3}}
\end{figure}

\begin{figure}
\centering
\includegraphics[angle=0,width=0.45 \textwidth]{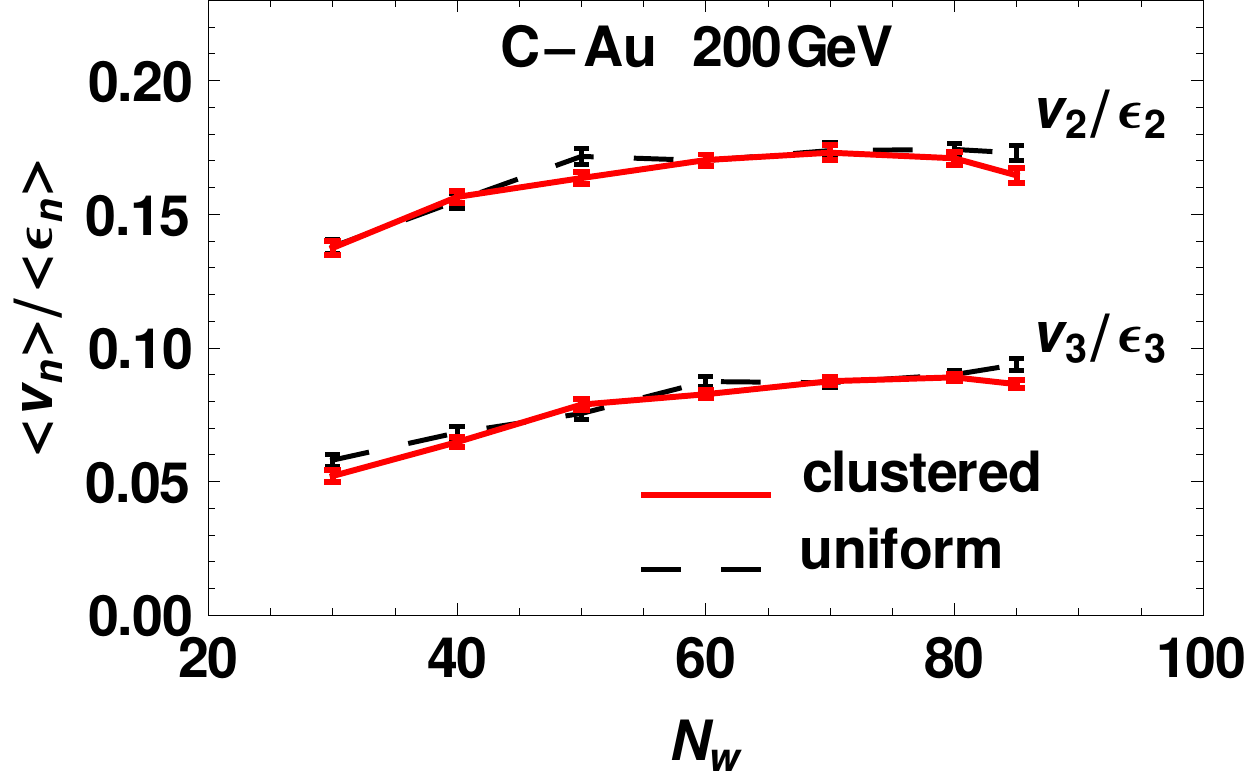}
\vspace{-3mm} 
\caption{(Color online) The hydrodynamic response coefficients for ellipticity 
and triangularity plotted as functions of the total number of wounded nucleons. 
BEC case, RHIC, 3+1 dimensional viscous hydrodynamics.  \label{fig:slope23}}
\end{figure}

\subsection{Flow coefficients \label{sec:flow}}

The eccentricity coefficients described in the previous Section are
not directly observable, as they correspond to the stage right
after one nucleus impinges on the other. In a hydrodynamic
approach this is just the initial stage, whereas what one measures are the
harmonic flow coefficients $v_n$, defined as the Fourier coefficients
of the azimuthal dependence of the particle spectra, namely
\begin{equation}
\frac{dN}{ d\phi} = \frac{N}{ 2 \pi}\left[ 1+2\sum_n v_n 
\cos\left[ n(\phi-\Psi_n)\right] \right] 
\label{eq:vn}
\end{equation}
(here we consider the 
$v_n$ coefficients integrated over the transverse momentum).

Realistic modeling of the flow coefficients requires advanced 
simulations of all stages of the reaction, from the early phase, 
through the intermediate hydrodynamics (for
reviews see, e.g.,~\cite{Heinz:2013th,Gale:2013da} and references therein)
or transport~\cite{Lin:2004en}, to hadronization at 
freeze-out (see, e.g.,~\cite{Florkowski:2010zz} for a review). Such 
event-by-event simulations have been carried out for numerous reactions, 
displaying collectivity even for such small system as in d-Au and p-Pb
collisions~\cite{Bozek:2011if,Adare:2013piz,Sickles:2013mua,%
Bozek:2013ska,Bzdak:2013zma,Qin:2013bha}.

A crucial finding for our method is that to a very good accuracy the
flow coefficients are proportional to the initial eccentricity
coefficients
\cite{Gardim:2011xv,Niemi:2012aj},
\begin{equation}
v_n = \kappa_n \epsilon_n, \;\; n=2,3.
\label{eq:linear}
\end{equation} 
The response coefficients $\kappa_n$ depend on the details of the
system and model (collision energy, multiplicity, viscosity of
quark-gluon plasma, initial time of collective evolution, freeze-out
temperature, feature of the applied ``afterburner''), yet the
linearity of Eq.~(\ref{eq:linear}) allows for model-independent
studies for certain quantities as shown in the following
Sections. These relations, realizing the shape-flow transmutation phenomenon, 
buttress quantitatively the naive
expectation of geometry-preserving features addressed in the
introduction.

We have carried out hydrodynamic simulations for the studied case 
of $^{12}$C-$^{197}$Au collisions at RHIC energies. We have used the
event-by-event 3+1 dimensional viscous hydrodynamics of 
Ref.~\cite{Bozek:2011ua}. At the freeze-out temperature of $150$~MeV hadrons 
are emitted following the
statistical hadronization model~\cite{Kisiel:2005hn,Chojnacki:2011hb}. The 
hydrodynamic 
evolution and particle emission at freeze-out include effects of the shear 
viscosity with $\eta/s=0.08$ and the bulk viscosity with $\zeta=0.04$ 
\cite{Bozek:2009dw}.
In Fig. \ref{fig:v2v3} we present the obtained estimates for the elliptic 
$v_2\{2\}$ and triangular $v_3\{2\}$  (all charged hadrons, 
$150$~MeV$<p_\perp<2$~GeV, $|\eta|<1$) 
flow coefficients for two cases: the 
clustered and the uniform initial $^{12}$C configurations. It is not possible to 
obtain directly accurate estimates
of the  higher-order cumulants $v_n\{4\}$ or $v_n\{6\}$  due to prohibitive 
requirements on the event-by-event statistics. 

Our simulations confirm the approximate linearity of Eq.~\ref{eq:linear}. The 
resulting 
response coefficients are presented in Fig.~\ref{fig:slope23}. We note that the 
response coefficients grow with $N_w$. The relative growth is quite strong, in 
particular $\kappa_3$ increases by 50\% from $N_w=30$ to $N_w=80$.

\subsection{Ratios of flow coefficients \label{sec:ratios}}

The above-mentioned increase of the response coefficient with multiplicity poses
an obstacle in qualitative analyses of the clusterization effect. Imagine that 
the experiment finds a growth of triangularity with $N_w$. A priori, without a
detail knowledge of the structure of the initial state and details of the 
dynamics, we cannot say how much of this growth should be attributed to 
intrinsic geometric deformation, and how much comes from the enhanced 
hydrodynamic response. One may avoid this difficulty by taking ratios of 
moments of event-by-event distributions of $\epsilon_n$ which are independent 
of $\kappa_n$. Thanks to the proportionality (\ref{eq:linear}), the same 
relations hold for the moments of $v_n$. Two popular choices with the low-order 
moments are the scaled standard deviation
\begin{eqnarray}
\frac{\sigma(\epsilon_n)}{\langle \epsilon_n \rangle} \simeq 
\frac{\sigma(v_n)}{\langle v_n \rangle}, \label{eq:ssd}
\end{eqnarray}
and the ratio of the four-particle and two-particle cumulant moments,
\begin{eqnarray}
\frac{\epsilon_n\{4\}}{\epsilon_n\{2\}} \simeq 
\frac{v_n\{4\}}{v_n\{2\}}. \label{eq:42}
\end{eqnarray}
Thus measurements of the above combinations of moments of $v_n$ provide 
information on analogous quantities for the eccentricities. Experimentally, one can access even moments of $v_n$, and  the ratio 
in Eq.~(\ref{eq:ssd}) must be estimated from $v_n\{2\}$ and $v_n\{4\}$ 
or from the reconstructed $v_n$ distribution.

\begin{figure}
\centering
\includegraphics[angle=0,width=0.46 \textwidth]{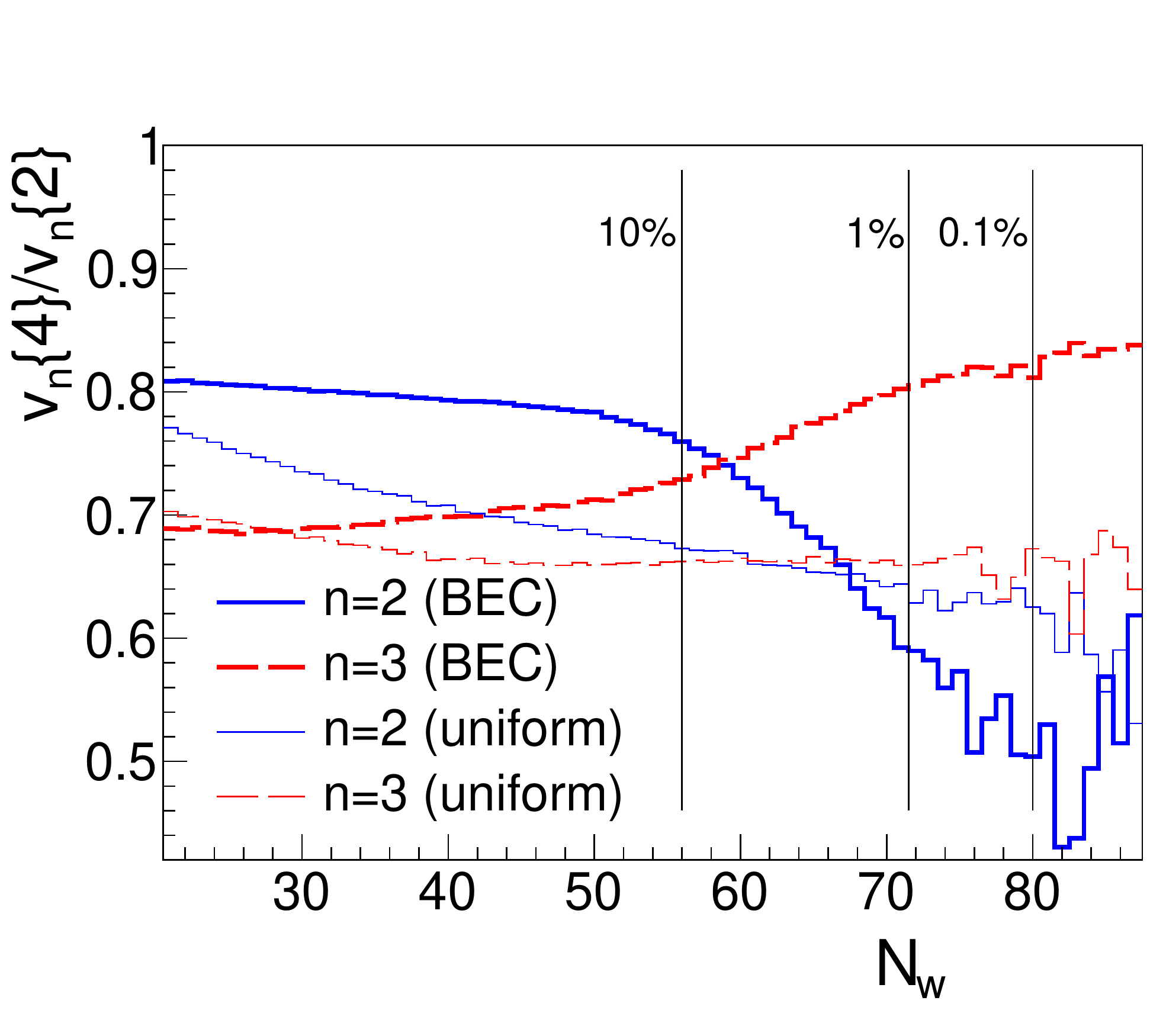}
\vspace{-3mm} 
\caption{(Color online) Ratios of four-particle to two-particle cumulants 
 plotted as functions of the total number of wounded 
nucleons. BEC case, RHIC. \label{fig:comb_ratios}}
\end{figure}

Predictions based on Eq.~(\ref{eq:ssd}) may be found already in 
Ref.~\cite{Broniowski:2013dia} (top-right Fig.~3 in that work), where 
${\sigma(\epsilon_n)}/{\langle \epsilon_n \rangle}$ grows for ellipticity and 
decreases for triangularity with $N_w$. This behavior reflects the interplay 
of the intrinsic geometry and statistical fluctuations. In this paper, 
following closely the analysis of Ref.~\cite{Bozek:2014cya}, we apply relation 
(\ref{eq:42}). The results of GLISSANDO simulations are shown in 
Fig.~\ref{fig:comb_ratios}. We note that for high multiplicity collisions the 
ratio ${\epsilon_n\{4\}}/{\epsilon_n\{2\}}$ significantly grows for 
triangularity and decreases for ellipticity. The geometric
triangularity increases for collisions with a larger number
of participants, corresponding to high multiplicity events.
On the other hand, the eccentricity due to fluctuations
of independent sources decreases with $N_w$, hence the opposite behavior.

We note that the change of behavior (stronger monotonicity) starts at $N_w$ 
corresponding to centrality of 10\%, thus occurs in the region easily 
accessible to experimental analyses.
We also see that the behavior for the 
clustered $^{12}$C (thick lines in Fig.~\ref{fig:comb_ratios}) is completely 
different from the case of the uniform structure (thin lines). 

The behavior shown in Fig.~\ref{fig:comb_ratios} is the key result of this 
work. It offers a signature sensitive to the intrinsic deformation that is 
straightforward to measure in ultra-relativistic heavy-ion collisions with 
standard techniques devoted to analysis of harmonic flow. 

One could ask at this point why we do need to resort to
Eq.~(\ref{eq:42}), rather than evaluate $v_n\{4\}$ directly from the
event-by-event hydrodynamic calculations. The reason is two-fold.
First, the statistics possible to achieve in such studies is
sufficient for the analysis of two-particle cumulants, but not
four-particle cumulants. Secondly, and more importantly, the
application of Eq.~(\ref{eq:42}) frees us from sensitivity to details
of the dynamical theory, which we do not know exactly. That way the
predictions for the ratios of the cumulant moments are more general
and model-independent.

\section{Further results \label{sec:results}}

\subsection{Dependence on the collision energy \label{sec:energy}}

\begin{figure}
\centering
\includegraphics[angle=0,width=0.46 \textwidth]{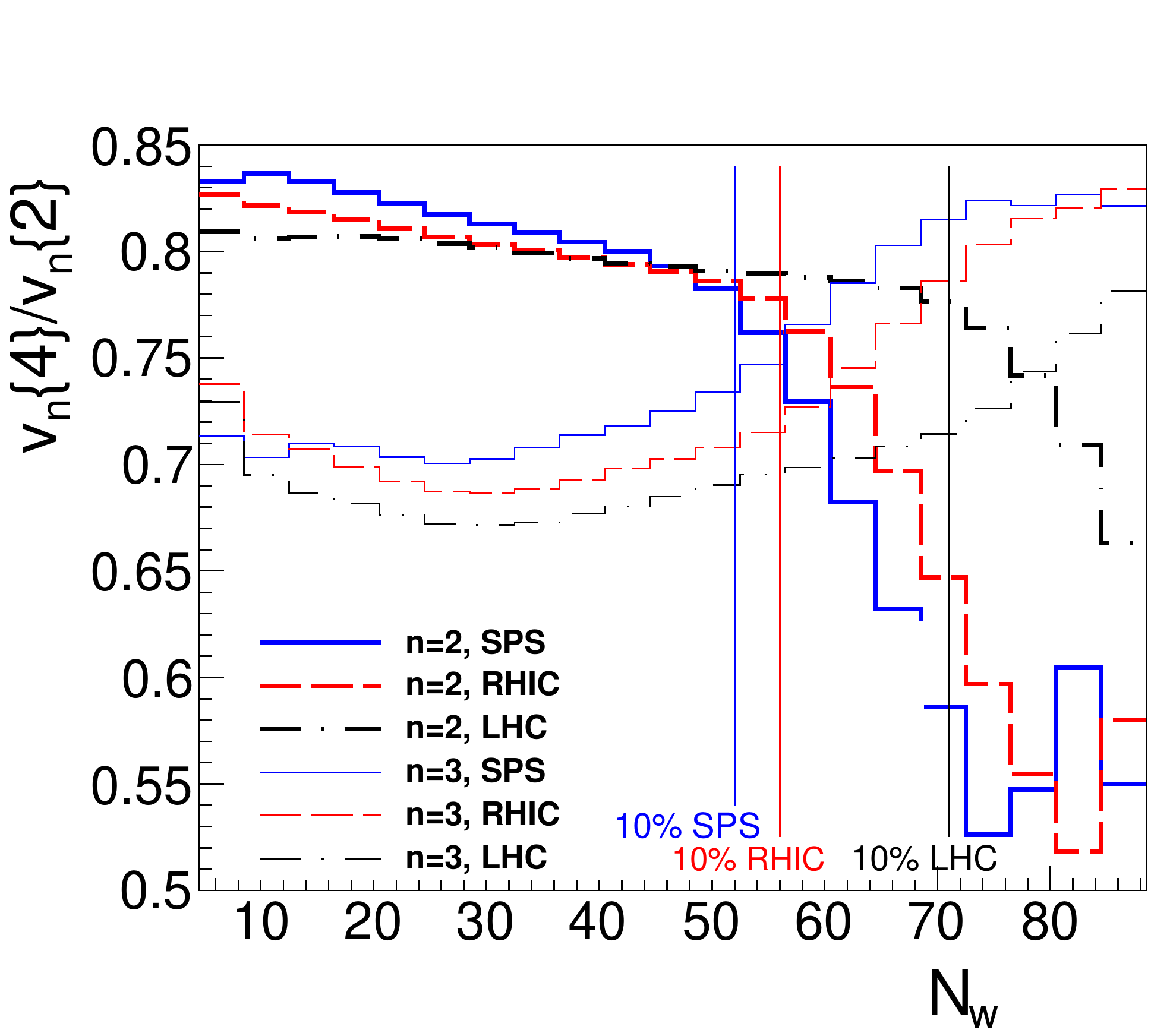}
\vspace{-3mm} 
\caption{(Color online) Comparison of 
$v_n\{4\}/v_n\{2\}$ for the SPS, RHIC, and LHC cases. BEC case. 
The vertical lines indicate the values of $N_w$ corresponding to centralities 
10\% for the three collision energies. Parameters are listed in 
Table~\ref{tab:par}.
\label{fig:energy}}
\end{figure}

In Fig.~\ref{fig:energy} we show the dependence of our predictions on the 
collision energy, according to the values in Table~\ref{tab:par}. We note that 
the qualitative predictions do not change with the collision energy, as the 
three sets of curves are similar, in particular when we take into account the 
fact that the values of centrality corresponding to a given $N_w$ depend on the 
energy via the value of $\sigma_{\rm NN}^{\rm inel}$.   

\subsection{Forward and backward rapidity \label{sec:rapidity}}

\begin{figure}
\centering
\includegraphics[angle=0,width=0.5 \textwidth]{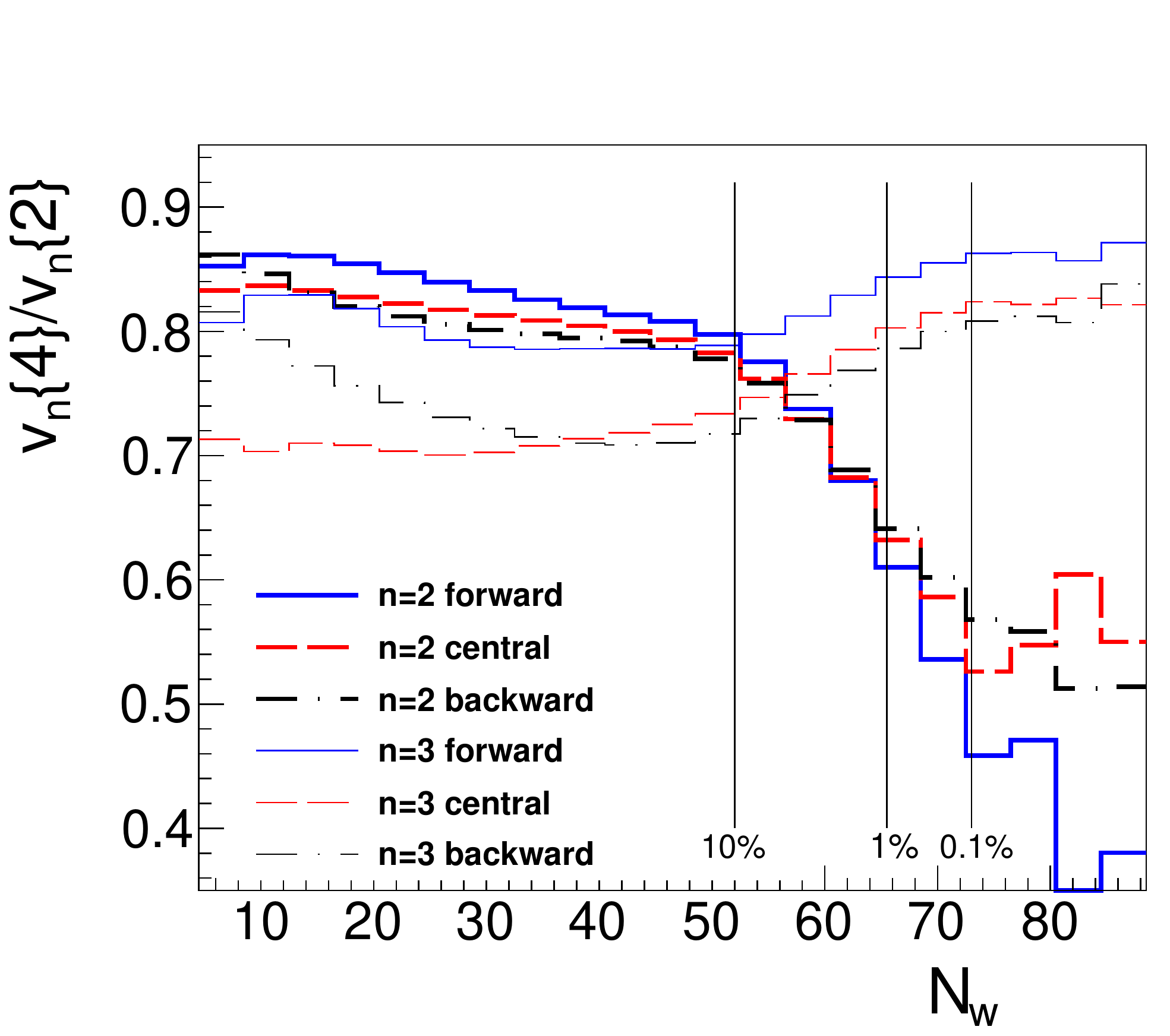}
\vspace{-3mm} 
\caption{(Color online) Ratios of the four-particle to two-particle cumulants 
 in forward, central, and backward rapidity regions, 
plotted as functions of the total number of the wounded 
nucleons. BEC case, SPS.\label{fig:fbc}}
\end{figure}

We may also ask the question on how much the predictions depend
on the rapidity window used in the experiment. This is of practical
significance, as in fixed-target experiments the detectors cover
rapidity away from the center. For the purpose of a simple estimate,
we use the model of Refs.~\cite{Bialas:2004su,Bozek:2010bi}, where the initial
density of the fireball in the space-time rapidity $\eta_\parallel$
and the transverse plane coordinates ($x, y$) is given by the form
\begin{eqnarray}
F(\eta_\parallel,x,y)&=&(1-a)[\rho_+(x,y) f_+(\eta_\parallel)
+ \rho_-(x,y) f_-(\eta_\parallel)] \nonumber \\
&+&a
\rho_{\rm bin}(x,y) \left [ f_+(\eta_\parallel) + f_-(\eta_\parallel) \right ], 
\label{eq:em}
\end{eqnarray}
where $\rho_\pm(x,y)$ is the density from the forward- and backward-going 
wounded nucleons, $\rho_{\rm bin}(x,y)$ is the binary collisions density. The 
rapidity profile functions $f_+(\eta_\parallel)$ and $f_-(\eta_\parallel)$ 
are given explicitly in Ref.~\cite{Bozek:2010bi}.

For our purpose it only matters that at mid-rapidity $f_\pm(0)=1/2$, 
hence $F(0)=(1-a) \frac{1}{2} (\rho_+ + \rho_-) + a \rho_{\rm bin}$, which is 
nothing else but the density of the mixed model used up to now to evaluate the 
fireball eccentricities. At very forward rapidities $\eta_+$ the value of 
$f_-(\eta_+)$ is negligible compared to $f_+(\eta_+)$, hence the relevant 
density of the fireball is $F(\eta_+)=f_+(\eta_+) \left [ (1-a) \rho_+ + 
a\rho_{\rm bin} \right ]$. Analogously, at large backward rapidities $\eta_-$ 
we have $F(\eta_-)=f_-(\eta_-) \left [ (1-a) \rho_- + 
a\rho_{\rm bin} \right ]$. 

The results of the calculation using central, forward (i.e., the
$^{12}$C hemisphere), and backward rapidity windows, with the source
density constructed according to the above-described prescription, is
shown in Fig.~\ref{fig:fbc}.  We note that the results do not alter
much with the choice of rapidity. We can see that the forward case has
the largest ratio for the triangularity, which follows from the fact
that it is more sensitive to the distribution of nucleons in the
clustered $^{12}$C. However, the monotonic behavior is not very
different, and there should not be much difference in the
results obtained at mid-rapidity in colliders and peripheral rapidity
in fixed-target experiments.

\subsection{Dependence on the model of the initial state \label{sec:depini}}

\begin{figure}
\centering
\includegraphics[angle=0,width=0.46 \textwidth]{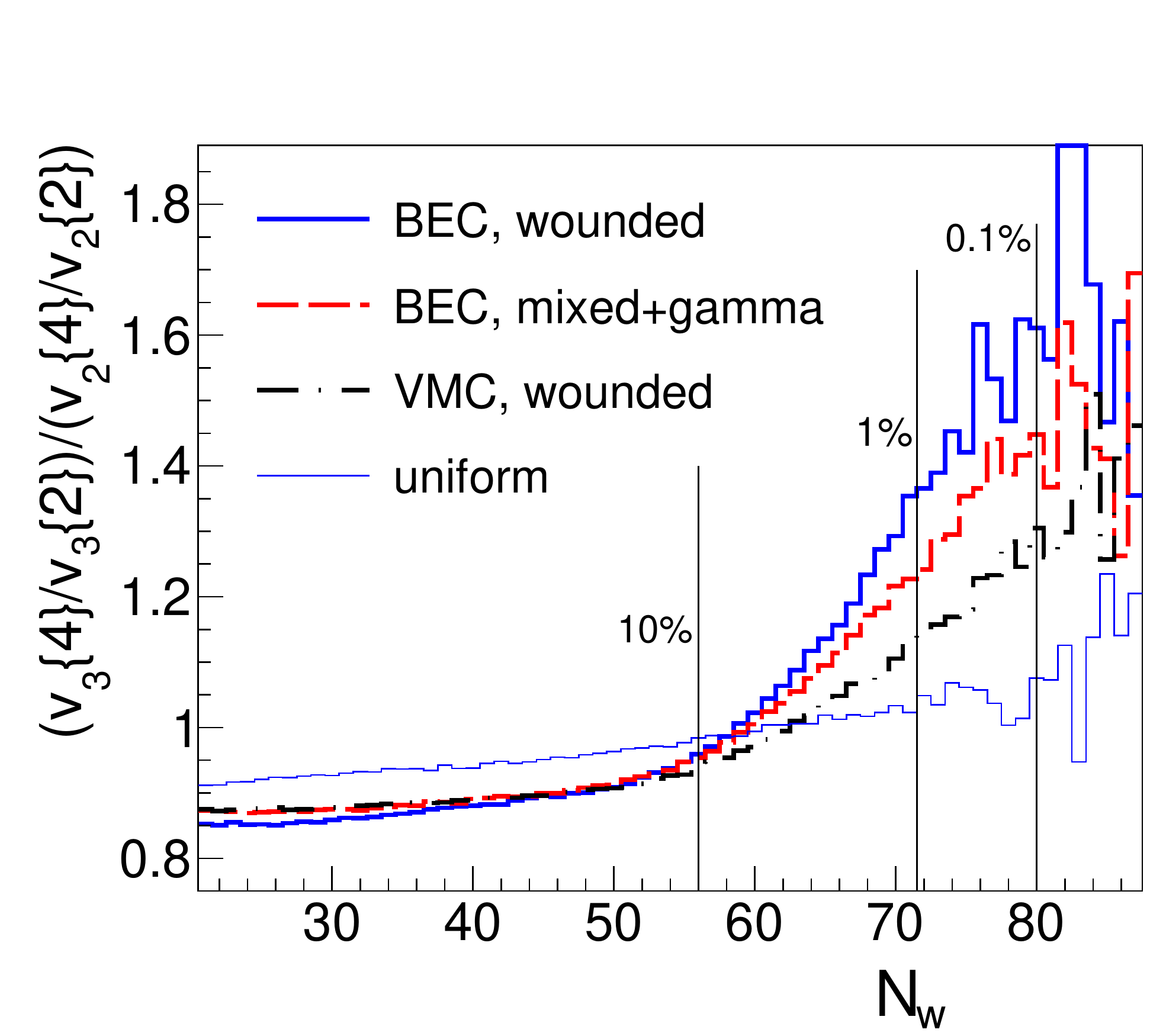}
\vspace{-3mm} 
\caption{(Color online) Comparison of the double ratio 
$(v_3\{4\}/v_3\{2\})/(v_2\{4\}/v_2\{2\})$ for various models, plotted as 
functions of the total number of wounded nucleons, RHIC. \label{fig:dr}}
\end{figure}

Up to now we have used the mixed model of the formation of the initial 
state, and the BEC distribution in $^{12}$C. In this subsection we study other 
variants of the Glauber approach, as well as the VMC distributions. The studied 
effects of clusterization depends to some extent on the models of the initial 
state, as they involve different amount of fluctuation. They also obviously 
depend on the configuration of the ground state of $^{12}$C. In 
Fig.~\ref{fig:dr} we present the double ratio 
$(v_3\{4\}/v_3\{2\})/(v_2\{4\}/v_2\{2\})$ for several cases. 

We notice the difference between the wounded nucleon model (less
fluctuations) and the mixed model with the overlaid gamma
distribution~\cite{Broniowski:2007nz} (significantly more
fluctuations). On the other hand, the result for the VMC distributions
is, as expected from the discussion of Sec.~\ref{sec:genc},
significantly weaker. Such a sensitivity is desired, as it in
principle allows for a quantitative discrimination of the $^{12}$C
wave functions from the studies of flow in ultra-relativistic
collisions.

\subsection{Deformed triangle \label{sec:deformed}}

\begin{figure}
\centering
\includegraphics[angle=0,width=0.46 \textwidth]{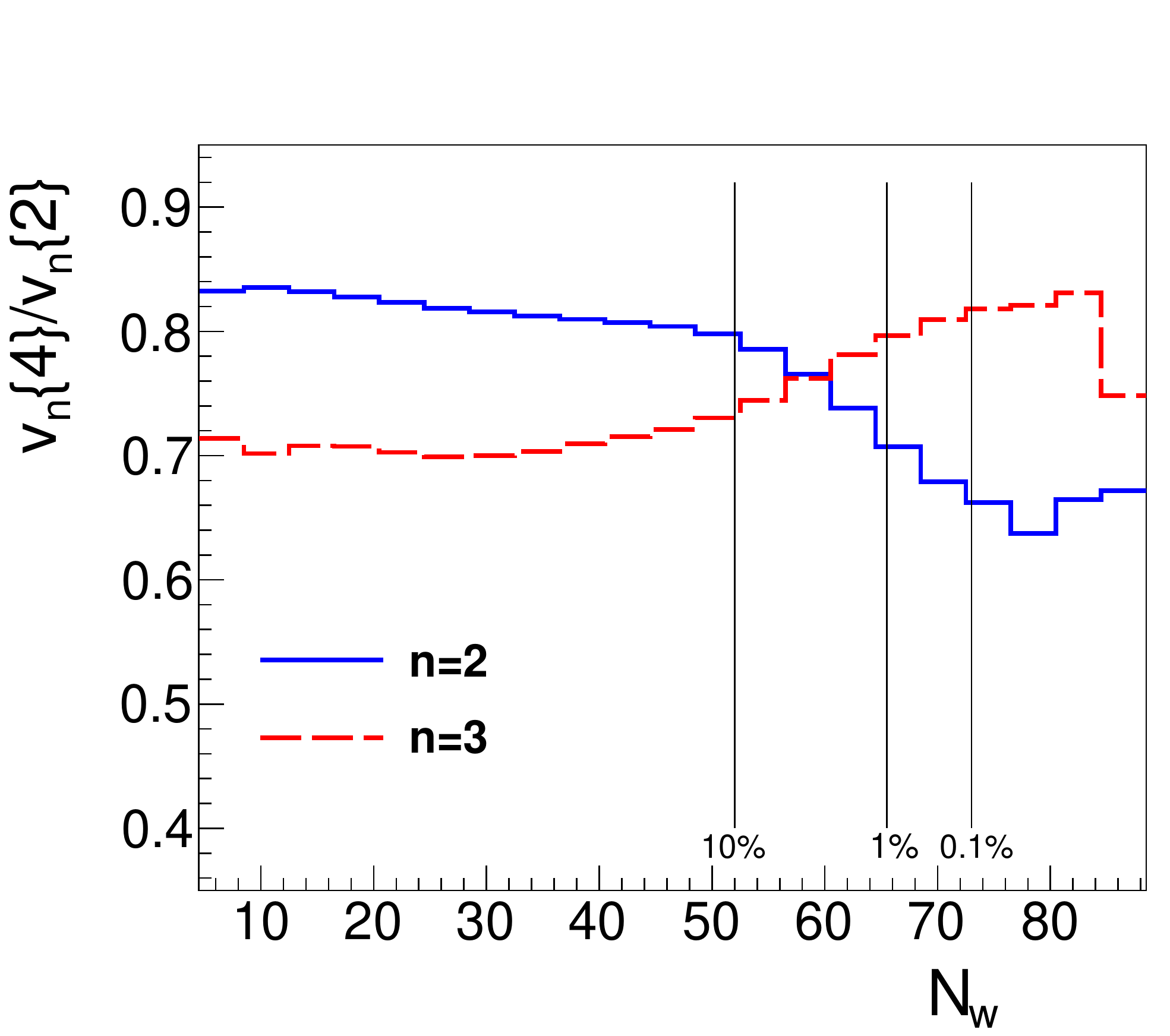}
\vspace{-3mm} 
\caption{(Color online) Predictions for
$v_n\{4\}/v_n\{2\}$    for the $^{12}$C distribution where the 
$\alpha$ clusters are arranged in a deformed triangle 
mimicking the FMD calculation of Ref.~\cite{Chernykh:2007zz}. SPS case. See 
text for 
details. \label{fig:deformed}}
\end{figure}

Finally, we study the case where the $^{12}$C nucleus is formed by placing the 
three $\alpha$ clusters in a deformed isosceles triangle, as results from the 
Fermionic Molecular Dynamics (FMD) studies of Ref.~\cite{Chernykh:2007zz}. For 
that purpose we take the ratio of the length of the edges to be 3/4, the value 
that can be read off from the right-most plot in Fig.~2 of 
Ref.~\cite{Chernykh:2007zz}.

The results of this calculation displayed in Fig.~\ref{fig:deformed} show that 
the qualitative behavior is the same as for the case of the equilateral 
triangle studied in the preceding sections. 

\section{Conclusions \label{sec:concl}}

We have pursued the idea that there is a geometry 
preserving principle operating in ultrarelativistic collisions, involving 
light nuclei impinging on heavy ions. The extremely high
energies make the interaction time so short that the existing granular
structures present in the ground-state nuclear wave function are effectively 
frozen.  The collision process realizes a
snapshot which may be recorded as a pattern in the collective flow emerging from
the abundantly produced particles in individual nucleon-nucleon
inelastic collisions. 

The presence of fluctuations due to the finite
number of particles as well the limitations imposed by the finite
nuclear size cause the geometric signals to be generically blurred
by random fluctuations. 
For that reason a careful analysis, as described in this paper, must be carried 
out.

We have addressed the question of intrinsic triangularity
structure of $^{12}$C motivated by the ancient idea that it has a
cluster structure with the three $\alpha$-particles sitting in the
corners of an equilateral triangle and the fact that the 12 nucleons 
induce a sufficiently large collectivity when colliding with heavy ions such
as $^{197}$Au or ${}^{208}$Pb. Other light isotopes in the region
$A \sim 10$ may undergo
a similar study as the one conducted here.

Furthermore, our analysis is insensitive to the hydrodynamic details by relying 
on the linear response of the harmonic
flow to the initial conditions. For that purpose, we examine the ratios of
cumulant moments, which are very well suited for our strategy. We find that 
visible
signals are expected in the dependence of these ratios on the number of wounded
nucleons, a feature which makes them particularly suitable for
experimental analysis. 

The fascinating possibility of catching
the $^{12}C$ nucleus as a triangle of $\alpha$ particles in ultrarelativistic
collisions would provide further evidence of Gamow's idea, taken from a
different angle as traditionally expected, and could be discerned
on experimental grounds. A positive answer would also generate further
confidence on the currently intricate theoretical approaches.

\begin{acknowledgments}
This research was supported by by the Polish National Science Centre,
grants DEC-2011/01/D/ST2/00772 and DEC-2012/06/A/ST2/00390, Spanish
DGI (grant FIS2011-24149) and Junta de Andaluc\'{\i}a (grant FQM225),
and PL-Grid infrastructure.
\end{acknowledgments}

\bibliography{from_hep,clusters,hydr}

\end{document}